Aalto University
School of Science
Bachelor's Programme in Science and Technology

# Nix: A Solution With Problems

**Bachelor's thesis**

**10 July 2026**

**Matias Zwinger**

Aalto-yliopisto
Perustieteiden korkeakoulu
Teknistieteellinen kandidaattiohjelma

KANDIDAATTITYÖN TIIVISTELMÄ

| **Tekijä** | Matias Zwinger |
|---|---|
| **Työn nimi** | Nix: A Solution With Problems |
| **Päiväys** | 10. heinäkuuta 2026 |
| **Sivumäärä** | 46 |
| **Pääaine** | Tietotekniikka |
| **Koodi** | SCI3027 |
| **Vastuuopettaja** | Lauri Savioja |
| **Työn ohjaaja** | Sergei Kozlukov |

Ohjelmistojen käyttöönotto (software deployment) on ohjelmistojen jakelua loppukäyttäjien koneille. Tätä varten tehdyt työkalut kärsivät monista ongelmista, jotka liittyvät esimerkiksi toistettavuuden puutteeseen ja riippuvuuksien ratkaisemiseen. Monet näistä ongelmista on onnistuneesti ratkaistu Nix-projektissa implementoidun puhtaasti funktionaalisen paketinhallintaparadigman avulla. Nix on Eelco Dolstran vuonna 2003 julkaisema ohjelmistoprojekti, jonka pääasiallinen komponentti on edellämainittua paradigmaa hyödyntävä paketinhallintatyökalu. Nix ei kuitenkaan ole onnistunut ratkaisemaan kaikkia tavanomaisten työkalujen ongelmia. Se on jopa luonut uusia, uniikkeja ongelmia. Tämän opinnäytetyön tavoitteena on kirjallisuuskatsauksen avulla kartoittaa Nixin nykyisen tutkimustyön tilanne ja ehdottaa Nixin yhteisölle suuntaa tulevaisuuden tutkimus- sekä kehitystyölle.

Opinnäytetyö on jaettu kolmeen osa-alueeseen. Ensimmäiseksi tarkastellaan ongelmia, joita ohjelmistojen käyttöönoton eri osa-alueilla on historiallisesti kohdattu, esimerkiksi toistettavuuden puute, riippuvuuksien ratkaiseminen ja luottamus. Neljä pääasiallista ohjelmistokategoriaa joita tutkitaan, ovat rakennusjärjestelmät, paketinhallintatyökalut, konfiguraationhallinta ja kehitysympäristöt. Kunkin kategorian suosittuja työkaluja käytetään esimerkkeinä ongelmien havainnollistamiseksi. Seuraavaksi esitellään Nix lyhyellä teknisellä katsauksella ja näytetään, millä tavalla se ratkaisee joitain aiemmin esitettyjä ongelmia.

Koska Nix on ensimmäinen suuri projekti, joka hyödyntää tätä verrattain uutta ja eksperimentaalista funktionaalista lähestymistapaa, se on kaukana täydellisestä ratkaisusta. Siitä syystä kolmanneksi analysoidaan Nixin aiheuttamia uusia ongelmia sekä vanhoja ongelmia, joita Nix ei ole toistaiseksi pystynyt ratkaisemaan, esimerkiksi luottamus sekä rakennusympäristön epäpuhtaus. Lisäksi käsitellään muutamia Nix-yhteisön esittämiä ratkaisuja näihin ongelmiin.

Nix syntyi alun perin puhtaasti akateemisena projektina ja se näkyy edelleen teknisen velan muodossa. Nix on kuitenkin tällä hetkellä paras kandidaatti ohjelmistojen käyttöönoton ongelmien ratkaisuun. Yhteisön tulisi alkuperäisen Nix-projektin parantelun ja kehityksen ohella myös tutkia vaihtoehtoja, kuten Snix- ja GNU Guix -projekteja. Vaikka Nix kärsiikin teknisestä velasta ja vaihtoehdot ovat vähissä sekä vielä kehityksensä alkupäässä, vaikuttaa puhtaasti funktionaalinen paradigma parhaimmalta tieltä tulevaisuuden kehitykselle.

Aalto University
School of Science
Bachelor's Programme in Science and Technology

BACHELOR'S THESIS
SUMMARY

| | |
|---|---|
| **Author** | Matias Zwinger |
| **Thesis title** | Nix: A Solution With Problems |
| **date** | 10 July 2026 |
| **No. of pages** | 46 |
| **Major** | Computer science |
| **Course code** | SCI3027 |
| **Main teacher** | Lauri Savioja |
| **Supervisor** | Sergei Kozlukov |

Software deployment suffers from numerous problems pertaining, for example, to reproducibility and dependency resolution. Many of these problems have been successfully solved by the purely functional approach to package management implemented in the Nix project. However, Nix does not solve all issues, and it does introduce some novel problems of its own. Therefore, the aim of this thesis is to conduct a literature review on the current state of research on Nix and to determine the direction of future research.

The first part of this paper explores the problems historically faced in different areas of software deployment, e.g., irreproducibility and dependency resolution issues. The main four categories of software deployment tools analyzed are build systems, package managers, configuration management, and development environments. Popular software from each category serve as case studies to illustrate the problems. The second part introduces Nix and explains the methods utilized to solve some of the problems introduced in the earlier part.

Because Nix is the first large project to utilize the purely functional approach, it is far from a perfect solution. Thus, the third part is dedicated to analyzing the new problems that Nix introduces, as well as old problems, which Nix has been unable to solve, such as trust and granular incremental builds. Furthermore, some proposed state of the art solutions put forth by the Nix community are discussed.

# 1. Introduction

> Software deployment is the problem of managing the distribution of software to end-user machines.
>
> — Eelco Dolstra [1]

Software deployment can be thought of as the "scaffolding" of software development, i.e., everything which is not designing, writing code, or testing. Under this definition, software deployment contains tools and software, such as build systems, package managers, configuration management, and software development tools, such as development environments. The main focus of research tends to be in software design, but software deployment is equally important for the end product as the coding itself.

Software deployment tools suffer from numerous problems pertaining, for example, to reproducibility and dependency resolution. Many of these issues have been solved by the Nix project with its purely functional approach.[1] It is, nonetheless, far from a perfect solution. Thus the aim of this thesis is to identify the extent of current research on Nix by conducting a state-of-the-art literature review.

The second section of this thesis enumerates some central problems in software deployment, while utilizing common tools as examples. The third section introduces the Nix software ecosystem with a short technical overview and explains the specific mechanisms it utilizes to solve the problems introduced in the previous section. The fourth section analyzes the novel problems created by Nix as well as those problems which Nix has been unable to solve. Moreover, it explores and discusses community-led state of the art solutions to these problems.

[1]The meaning of functional in this context is explained in Section A2.

# 2. The Current Landscape

The following sections will introduce some mainstream software deployment tools and demonstrate their shortcomings which have been largely mitigated by the Purely Functional model.

## 2.1. Build Systems

Build systems or build tools automate the process of compiling, linking, and sometimes installing[2] software. A build tool can be a simple Bash or Perl script which runs commands in sequence or complex software with integrated test suites. Most build systems work by defining build targets, and the intermediate steps necessary to reach that target. For example, if `main.c` links a function defined in `helper.c`, two object files, e.g., `main.o` and `helper.o`, are required to compile the executable `main`. In this case, the target would require two intermediate steps, one for generating each object file.

One of the earliest and most widely adopted build tools is Make [2]. it utilizes the modification times of the source code to minimize the amount of building required to reach a certain target: It builds `foo.o` only if the file does not exist or one of the input files of that step has a newer timestamp than `foo.o`. For a long time Make has been the go-to build tool, but due to the rising complexity of software projects, more sophisticated tools have been developed [3].

### 2.1.1. General Problems

*Unnecessary rebuilds* may occur if the modification time of a file is changed, but its contents remain the same, or if how the build system tracks changes does not align with the semantics of the language. For example, a comment changes the contents of the file but not the build output.

*Stale artifacts* can appear in the cache if the build system does not model some dependencies. For example, if `main.c` includes `helper.h`, Make does not know about this dependency because it does not analyze the source files. When `helper.h` is modified but `main.c` is not, any artifacts produced from `main.c` are stale. Stale artifacts can also result from changing the compiler or from upgrading a system library on which the program depends. In all cases, it is caused by a lack of dependency modeling by Make.

*Remote caching* is the ability of build systems to reuse build artifacts built on other machines [4]. Reusing build artifacts from earlier local builds works on Make, but remote caching is not supported. It is practical in situations where the local repository is no longer up to date, so building the project would necessitate rebuilding many targets, and someone else has already built the newest version.

[2]Moving build outputs (artifacts) to the end locations in the filesystem.

### 2.1.2. Reproducibility Problems

The produced output may vary if it is built on a different computer or even if it is built at a different time. Consequently, irreproducibility is an issue that has gained traction recently in many scientific fields, where results that depend on software need to be reproducible [5]. Some common causes of irreproducibility are listed by Lamb and Zacchiroli [6]. They can be sorted into build environment and language non-determinism.

**Build Environment Non-Determinism** is caused by the operating system, filesystem, hardware, etc. affecting the build output.

*Build timestamps* are timestamps generated at build time which are embedded in build results, for example, by the C `__DATE__` preprocessor macro.

*Build paths* are build directory paths embedded in the build results, for example, generated with the C `__FILE__` preprocessor macro. Other, more manageable problems are `RUNPATH` entries in executable and linkable format (ELF) binaries, which store the run-time search paths for libraries. These libraries may be located in different paths in different environments.

*Filesystem ordering* is the order in which filesystem objects (FSOs) such as files and directories are enumerated in. For example, the `readdir(3)` syscall does not specify an ordering for its results. If a build tool or compiler processes files in a random order, it may affect the build output.

*Archive metadata* is metadata saved in archive files. Some common archive formats such as ZIP and TAR store timestamps and file ownership information. This means that building with a different user ID (UID) or at a different time results in different outputs.

*Parallelism* in the build tool may result in non-determinism if the completion order of tasks is somehow encoded in the build artifact. For example, tasks A and B may run in 5 seconds and 10 seconds respectively on one CPU, but another CPU may need only 3 seconds to run B, and thus the order is reversed.

**Language Non-Determinism** is caused by the programming language itself.[3]

*Data structure ordering* may cause non-determinism when the language stores data structures in a random configuration, or iterates over data structures in a random order. For example, in many languages, iterating over hash tables yields items in a random order. If the result of such an enumeration of items is utilized anywhere in the build process,[4] the artifacts differ for every build.

[3]This usually means the compiler of the language, unless it is an interpreted language such as Python.

*Padding* is another source of data structure randomness. Some languages use random data as padding or to fill unused fields in data structures. Obviously, this data should never be read, but as most languages are not memory safe, it might still happen and cause non-deterministic build outputs.

## 2.2. Package Management

A package manager is a piece of software for installing and managing other software and their dependencies.[5] This may sound trivial, but dependency resolution can become difficult in many cases.

Traditional binary package managers such as dpkg [8], RPM [9], or Pacman [10] do this by downloading pre-compiled binaries from servers and moving these files to the correct locations on the system. In contrast, source-based package managers such as BSD Ports [11] or Portage [12] instead copy the whole source code of the package to the system and produce the binaries locally, which necessitates that they are also a build tool at the same time.

### 2.2.1. Dependency Issues

The most common problems in package management stem from dependency resolution. Two scenarios where this issue is apparent are presented below.

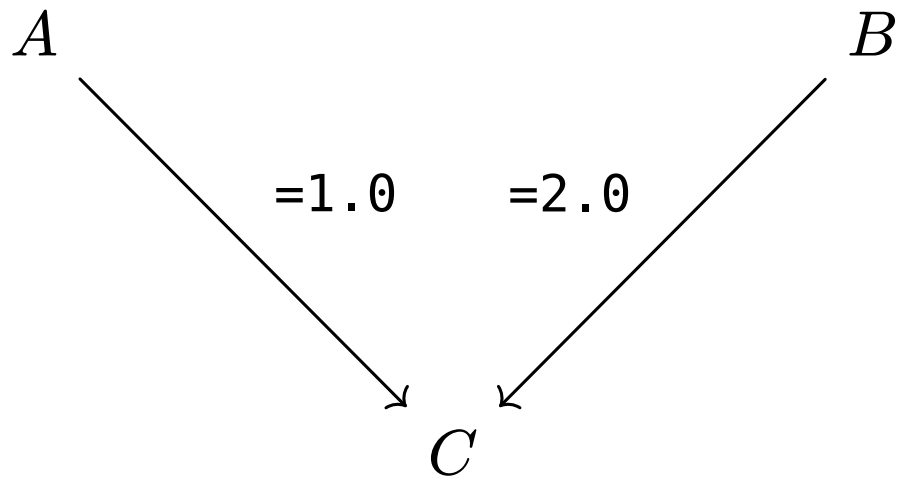

Figure 1: Conflicting dependencies. Arrows denote dependency relations, numbers denote version constraints.

*Conflicting dependencies* happen when packages A and B both depend on different versions of library C. A visual representation is presented in Figure 1.

[4]This can happen if a part of the build toolchain is written in any language exhibiting this behavior.
[5]Note that there exist package managers which do not perform dependency resolution, most notably Slackware's pkgtools [7].

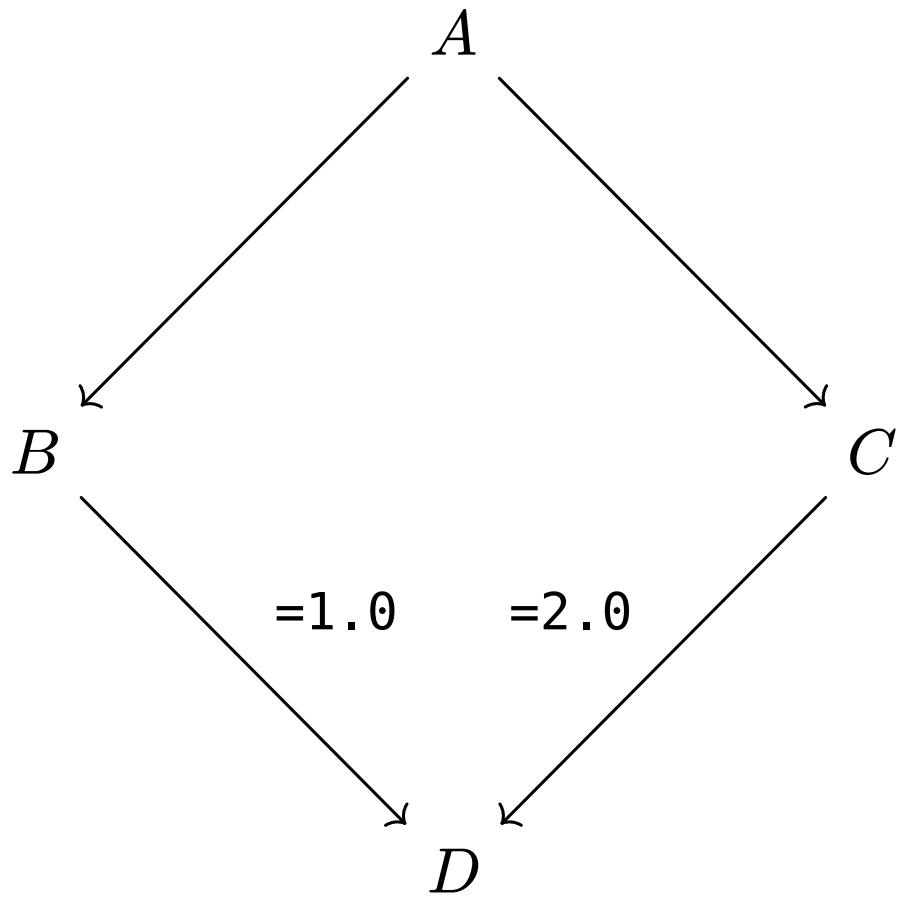

Figure 2: Diamond dependency relation. Arrows denote dependency relations. Numbers denote version constraints.

*Diamond dependencies* are a more complex case of conflicting dependencies. Package A depends on libraries B and C, which both depend on library D. However, B depends on D 1.0, and C depends on D 2.0. Since only one version of library D can be installed at a given time, it becomes impossible to install A. A visualization of this situation is presented in Figure 2.

#### 2.2.2. Trust

All intermediaries in software distribution are a possible attack vector [13], including the package repository. In binary package managers, the end-user needs to trust the owner of the package repository. The binaries served on the server may have been built by the repository owner with malicious code injected into them.

## 2.3. Configuration Management

Configuration management is a process for maintaining a desired state of IT systems and components [14]. There are two traditionally used ways to achieve this goal, convergent configuration management systems (CCMSs) and containerization [15].

#### 2.3.1. Convergent Configuration Management Systems

The term "convergence" in relation to configuration management is introduced by [16]. A CCMS tool works by modifying the target system(s) to match the desired state [15]. The actual state of the system approaches the target state as a function of time [16]. The desired state is described by administrators with a domain specific language (DSL), which the tool reads. In addition to being "recipes" describing the desired state, the configuration also serves as documentation, which describes the (approximate) state of the target system.

Ansible is one of the most prevalent CCMS tools and the current industry standard [17]. An Ansible configuration consists of YAML or INI files describing the inventory, i.e., a set of target hosts and playbooks which consist of individual tasks, i.e., provisioning operations [17]. Playbooks can be grouped further into roles. The configuration is stored on the developer's machine, and Ansible commands run different roles and playbooks [17]. On the implementational level Ansible works by copying Python scripts called "modules" to the target host over Secure Shell (SSH) and running them [17], [18].

Ansible and other CCMS tools work with the same central principle: The user writes a target state in a declarative language and the tool performs imperative tasks in order to change the state of the target system to match the description. This means that any manual changes on the system will cause the state veer off the target state. Take as an example an Ansible playbook which modifies the SSH config of a system to disallow password logins. If another administrator now logs into the server and changes the line back to allow them, the system does not match the state documented in the playbook.

Another issue arises when removing packages from the configuration. Take as an example a configuration which specifies the package `foo` to be installed. When the system is provisioned using the tool for the first time, `foo` is installed. Later, the administrators decide `foo` is no longer needed, and they remove the entry from the configuration which installs `foo`. However, when applying this configuration to the target(s), `foo` is not uninstalled. Instead, a temporary modification to the configuration has to be made which explicitly uninstalls `foo` from the system, after which the configuration can be applied [15]. This happens because, although the DSL is declarative, it actually describes a set of imperative steps which are executed on the host(s). For this reason rollbacks to previous configurations are hard and require manual intervention.

Two hosts with the same target state may end up with different versions of packages if they are provisioned at different times. Administrators can practice version pinning if the target package manager host(s) allows it, but it is not enforced by default.

These tools also do not guarantee idempotence. In configuration management idempotence means that running a task multiple times in a row should produce the same result as just running it once. While many modules in Ansible are designed to be idempotent, it is not guaranteed that they are [19].

### 2.3.2. Containers

Containers are isolated user-spaces which share the kernel with the host operating system [20]. This allows for lightweight isolated environments with less overhead than virtual machines. The intended use of containerization is for process and service management and isolation; However, the features can also be used for configuration management.

Docker is one of the most widely used containerization tools. It works by creating containers from immutable *images* [21]. Images are created from a base image, which acts as the first *layer*. The base image can be e.g., a full Ubuntu system or an empty filesystem. Each consecutive modification is its own layer which builds upon the previous layer. This allows for some amount of deduplication, as layers can be reused. In practice this deduplication is not always optimal, since containers are ran using all kinds of different base images. There is also a limit to how many layers an image can have [22], so the granularity of deduplication is limited. Furthermore a layer is modified and the image is rebuilt, all layers which build upon the modified layer need to be rebuilt too [15].

However, the lack of version pinning is solved: Images have inbuilt version pinning, since they are immutable. Once an image is created, each derived container will have identical filesystem contents. Furthermore, rollbacks become easy, as the previous configurations are still present in the form of images.

Containers alone are not sufficient for building congruent systems. They are only useful in configuration management insofar as they make it easy to specify an immutable image as a starting state, and make rolling back to that state easy.

## 2.4. Development Environments

Development environments, or dev envs for short, are shell environments which expose the necessary packages and libraries to develop software. This could include compilers and libraries but also linters and language servers. They work by modifying environment variables, most notably `PATH` and `LD_LIBRARY_PATH`. Virtual environments can be manually entered with a script or triggered when entering a certain directory [23]. Packages installed in a development environment are isolated from the rest of the system, for example, by installing them in a separate, local directory [24].

The incentive for developers to utilize dev envs is that sometimes projects require different versions of the same package. In essence, dev envs partially circumvent the issues described in Section 2.2.1 by avoiding inter-project conflicts entirely.

Dev envs separate package installations on a per-project basis, so issues faced by package managers are still possible within a single project. In other words, dev envs only avoid inter-, but not intra-project conflicts.

A corollary of installing packages in separate namespaces for each project is that disk space is not utilized optimally.[6] Two projects sharing an identical dependency are required to install separate instances of the same files, since they are in separate filesystem paths, leading to duplication.

[6]Except if the filesystem has built-in deduplication features.

# 3. Nix Solves This

The Nix package manager and build tool started as a research project by Eelco Dolstra in 2002 in the form of Maak [25].[7] Nix solves many of the problems mentioned in Section 2 by employing a host of different techniques presented in this section.

## 3.1. The Structure of the Nix Project

Nix is not a single piece of software such as Make or Maak. Instead, it is an ecosystem of various projects. Here are the four most important components of the Nix project:

(1) The Nix package manager,[8] which also includes the build tool, as well as all the Nix command line tools and the Nix daemon [26], [27];[9,10]

(2) The Nix DSL, which Nix packages are specified in;[11]

(3) nixpkgs, the largest[12] collection of Nix packages [29];

(4) The NixOS Linux distribution [30].

## 3.2. Reproducibility

### 3.2.1. The Nix Language

Nix advertises itself as reproducible[13] and declarative [30], which in large parts can be attributed to the Nix language.[14] The Nix package manager requires every package to be defined using this purely functional, dynamically typed DSL. The largest collection of these package definitions is called nixpkgs, the official Nix package repository [29].

---

[7] "Maak" translates to "to make" in Dutch, an homage to the Make build tool which was mentioned in Section 2.1.

[8] The original Nix package manager is commonly called NixCPP to distinguish it from forks and rewrites. CPP stands for the C++ programming language, which NixCPP is written in [26].

[9] In single-user mode the store is owned by the user and modified directly by that user using the Nix command line tools. The Nix daemon is necessary only in multi-user installations of Nix [1], [27]. A multi-user system will be assumed for the rest of this thesis if not otherwise specified.

[10] Section A1 shows a concept map of the Nix package manager as it appears in the Nix Reference Manual [27].

[11] Technically the Nix DSL is defined in the source code of NixCPP, but it makes sense to treat it as a separate concept in the same sense as a language and its compiler are treated as separate concepts.

[12] There exist some smaller package collections too, most notably the Nix User Repository [28].

[13] In reality, Nix does guarantee reproducibility, see Section 4.7. A more fitting description would be *repeatable*, which means that Nix builds will always produce a correct output, e.g., a runnable binary, even though separate builds may not be bitwise reproducible. I.e., a realization of a derivation will succeed if it has succeeded once. Derivations will be discussed in depth in Section 3.2.3.

[14] A proper introduction to the language syntax is outside the scope of this thesis, but there are a number of tutorials on the internet, including [31].

At the heart of the Nix package manager is the standard environment called stdenv. It describes a basic POSIX build environment, containing GCC, Bash, and other fundamental POSIX tools [1].

```
{
  stdenv,
  fetchurl,
}:

stdenv.mkDerivation {
  pname = "hello";
  version = "2.12.2";

  src = fetchurl {
    url = "mirror://gnu/hello/hello-${finalAttrs.version}.tar.gz";
    hash = "sha256-WpqZbcKSzCTc9BHO6H6S9qrluNE72caBm0x6nc4IGKs=";
  };
}
```

Listing 1: A simplified Nix expression for GNU Hello

stdenv automatically runs `./configure`, `make`, and `make install` in the directory downloaded from `src`. These are the three main phases, `configurePhase`, `buildPhase`, and `installPhase`. There are also other phases, such as `unpackPhase` (unpacking the compressed source), `checkPhase` (for running tests), and `fixupPhase` (post-processing of the installed files). The Bash scripts ran in these phases can be modified by specifying them in the body of the `mkDerivation` function, for example `buildPhase = "make --foo --bar"`. Build- and runtime dependencies can also be supplied as arguments to the `stdenv` function [32]. Because Nix forces users to model dependencies like this, stale artifacts are avoided.

### 3.2.2. Sandboxing

Nix requires build inputs to be stated explicitly, and it tries to enforce this requirement through sandboxing. This means that Nix builds are isolated from the rest of the system;[15,16] All build environments start from a uniform, clean state and only the necessary, explicitly stated components are made accessible inside it. They cannot access files other than those of dependencies or the source code. Even `/proc` and `/dev` are replaced with private versions [27]. The sandbox also creates private mount, network, IPC, and UTS namespaces for each build to further increase determinism [27].[17] In the sandbox environment variables are not inherited from the caller or the Nix process user.

[15] Unless impure evaluation is enabled.

[16] Sandboxing is enabled by default only on Linux systems.

[17] An exception to this are fetchers, which are not placed in a private network namespace as they need to be able to download files [27].

Instead, they must be explicitly defined. `PATH` is initially set to `/path-not-set` and `HOME` is set to the aptly named `/homeless-shelter` [1].

These measures partially solve the build environment non-determinism listed in Section 2.1.2 by enforcing an interface which restricts access to undeclared inputs. However, while very good, this restriction is not perfect, as will be seen in Section 4.7.

### 3.2.3. Derivations and the Nix Store

The function used to specify the package in Listing 1 is called `mkDerivation` and not `mkPackage` because Nix is a purely functional language and as such is unable to actually interact with the outside world and build something. Instead, the Nix expression is evaluated into a *derivation* in a process called instantiation. Derivations contain structured data represented in the ATerm format [33] and end in the `.drv` file extension. The Nix daemon can read a derivation and build it, which yields the final package. The derivation includes, i.a., the builder and its args,[18] environment variables, and system arch [1]. Most importantly, the derivation also contains paths to the derivations and realized versions of the dependencies. The set of transitive dependencies of a derivation is called the *closure* of a derivation, which is explained in more detail in Section A6. To summarize, a derivation is a specification of an execution of `execve` [34].

Both derivations and realized packages are located in a special directory called the Nix store, which is usually located in `/nix/store`. All components in the store have a hash as part of their path, for example `/nix/store/lz9gfg6iybsh0hiignpk55w99a3bj4vb-hello-2.12.1`. The path of a derivation file is called the `drvPath`, and the path of a build output is called the `outPath`. How store paths are constructed is explained in Section A3. Note that [1] describes two different Nix store architectures, the input-addressed (IA)-store and the content-addressed (CA)-store. What is described here is the IA approach, and all of Section 3 assumes IA. CA will be explored later in Section 4.2.1.

Before build results are moved to the Nix store they are *canonicalized*; All files are set to uniform permission levels[19] and timestamps (Unix time 0) [1]. Files which are not regular files, directories, or symlinks are not allowed and will result in an error [1]. This is done to further reduce possible sources of randomness, such as build times.

The hash for a store object of a build result is calculated over all the inputs specified in the derivation, which includes, i.a., the buildtime environment variables, store paths of the dependencies, builder script, and source code of the component. Therefore any change to an input component, the source code, or the build instructions will result in a different store path [1]. Hashing allows Nix to host multiple parallel versions of the same software without namespace conflicts, solving the dependency issues in Section 2.2.1. Another problem solved by hashing are unnecessary rebuilds. If neither source code nor build instructions change (i.e., unaltered inputs), Nix reuses the existing build result from the store. However, if comments are added to the source code, Nix treats the source code as changed and triggers a rebuild.

[18]The builder and its args can be e.g., a compiler and build options passed to it. It can be also as simple as a bash script which unpacks fonts into the right directory structure. Although it is called a builder, it does not have to build software in the way a compiler does.

[19]Permissions are set to `0444` for files and `0555` for directories and executables, and special permissions such as SUID and SGID are removed [1].

Because storing components such as binaries and libraries in the store necessitates breaking the filesystem hierarchy standard (FHS), Nix exposes the installed binaries to the user by creating a user environment. This user environment is a store object which contains symlinks to binaries the user is configured to access. The store path of the user environment is then added to the `$PATH` of the user [1].

### 3.2.4. Substitution

When the Nix daemon realizes a store path, it checks whether a component with that exact store path exists on a substituter before starting a local build. A substituter is a mechanism which provides store objects directly without the need to build them, i.e., essentially a binary cache [1]. The most common substituter type is a public server which exposes its Nix store to the internet [1]. The NixOS project runs a public substituter using their own Hydra software at https://cache.nixos.org. Fresh Nix installations are automatically configured to use this substituter [35].

The substituter mechanism allows Nix to be a source and binary package manager at the same time; Most packages will be substituted directly for their binary outputs by a substitutor like Hydra, but compilation from source is not treated as a second-class citizen.

## 3.3. NixOS

The Nix store can be used to store any filesystem object; Nothing says that only binary packages (normally stored in `/bin`) and libraries (normally stored in `/lib`) can be put in the Nix store. Any process which results in a file tree, i.e., filesystem object, can be stored as a component in the Nix store. This means placing operating system components like the kernel and initial ramdisk (`/boot`) and config files (`/etc`) into the store is allowed. This approach is realized in NixOS, a Linux distribution which not only uses Nix as its primary package manager, but stores its own components in the Nix store too [1], [36].[20]

A NixOS system configuration consists of configuration files written in the Nix language. The nixpkgs repository contains ready-made *modules* which generate system components, such as configuration files or systemd services. They expose *options* which can be modified by the user in the NixOS configuration. Of course users can also write and use their own modules [36].

[20] In reality, some parts of the bootloader cannot be used from the store, since the store has to be mounted by the initrd. See Section A9 for more information on the NixOS boot process.

```
services.openssh = {
  enable = true;
  openFirewall = false;
  settings = {
    PasswordAuthentication = false;
    KbdInteractiveAuthentication = false;
    PermitRootLogin = "without-password";
  };
};
```

Listing 2: NixOS configuration to enable an OpenSSH server. This the sshd NixOS module from nixpkgs.

```
options = {

   services.openssh = {

     enable = lib.mkOption {
       type = lib.types.bool;
       default = false;
       description = ''
         Whether to enable the OpenSSH secure shell daemon, which
         allows secure remote logins.
       '';
     };

     package = lib.mkOption {
       type = lib.types.package;
       default = config.programs.ssh.package;
       defaultText = lib.literalExpression "programs.ssh.package";
       description = "OpenSSH package to use for sshd.";
     };

# [...]

};
```

Listing 3: Part of the sshd NixOS module from nixpkgs.

```nix
users.users.sshd = {
  isSystemUser = true;
  group = "sshd";
  description = "SSH privilege separation user";
};
users.groups.sshd = { };

# [...]

environment.etc =
  authKeysFiles
  // authPrincipalsFiles
  // {
    "ssh/moduli".source = cfg.moduliFile;
    "ssh/sshd_config".source = sshconf;
  };

systemd.tmpfiles.settings."ssh-root-provision" = {
  "/root"."d-" = {
    user = "root";
    group = ":root";
    mode = ":700";
  };
  "/root/.ssh"."d-" = {
    user = "root";
    group = ":root";
    mode = ":700";
  };
  "/root/.ssh/authorized_keys"."f^" = {
    user = "root";
    group = ":root";
    mode = ":600";
    argument = "ssh.authorized_keys.root";
  };
};
```

Listing 4: Part of the sshd NixOS module. The sshd module uses other modules, such as the environment module and the systemd module.

Listing 2 shows how a module defined elsewhere is used in a NixOS configuration. In this case it is the sshd module from nixpkgs. Listing 3 is a part of the sshd module file which defines the options which are set by the user in Listing 2. Listing 4 is another part of the module, defining the sshd user, configuration file locations, and systemd tmpfile settings. The content of the configuration files generated by the module is based on the options defined in Listing 3 and specified in Listing 2.

When a NixOS configuration is built, it creates a generation — a closure of the system configuration — which includes all configuration files, programs, etc. as defined by the configuration. Each generation resides in `/nix/var/nix/profiles`, with the booted system being `/nix/var/nix/profiles/system`. Additionally, all profiles are symlinked to `/nix/var/nix/gcroots` to prevent the garbage collector from removing them.[21] In addition, the ephemeral symlinks `/run/booted-system` and `/run/current-system` point to the original booted system and current running configuration respectively [36].[22] Binaries meant to be exposed to the user are located in `/run/current-system/sw/bin`, which is added to `$PATH` [35]. Since switching generations only involves updating the symlink at `/run/current-system` to point to the new generation, this operation is atomic, which means that upgrading a NixOS system is an atomic operation [36].[23]

The Nix store does not delete unused generations unless manually instructed, so old generations remain bootable. This fact is leveraged in NixOS by showing old generations which still exist in the store as bootable entries in the bootloader, allowing easy rollbacks in case a system upgrade breaks something [36]. Another benefit of having all software in the Nix store and using environment variables to control what is exposed is that each user can have a different environment with different packages, libraries, and configuration files [27].

NixOS directly solves the issues described in Section 2.3 by getting rid of the imperative paradigm. Since the state of a NixOS system is contained in the Nix store, and the store can only be modified by the Nix daemon,[24,25] the system is immutable. This makes NixOS into what [16] calls a congruent configuration system. It does not try to move the hosts towards the target state, instead keeps them in complete compliance with it. The lack of idempotence is also solved, as building a system twice from the same configuration results in the exact same output (i.e., store hash), rendering its redeployment as unnecessary. Version pinning is solved too, as the NixOS configuration can be easily pinned to a certain nixpkgs version.

When compared to Docker or other containerization tools, Nix solves the issues of reproducibility, version pinning, deduplication, and rollbacks without introducing the notion of namespaces in the first place. The functional approach of Nix is compatible with and amplifies the benefits of containers, although jobs that can be completed with a simple Nix shell should probably be solved without containers. However, there are many legitimate use cases for containers, but even then Nix is more efficient than Docker at building images [37], [38].

[21]This action makes all profiles gcroots, see Section A5 for a more extensive explanation.

[22]This separation is necessary because the Kernel cannot be swapped without a reboot (unless kexec is used). Both point to the same location if no generation switch has occurred.

[23]Yet again, the bootloader is an exception to this. Since the main Nix store is not mounted before stage 2 (see Section A9), the contents of the bootloader cannot be symlinks from the main store, and hence the updating the contents of `/boot` is not atomic.

[24]That is, by modifying the system configuration and rebuilding the system.

[25]Again, assuming a multi-user installation.

## 3.4. Nix Shell

Nix shells are ephemeral environments which a user can open shells in on demand, e.g., when entering a specific directory. Since entering a Nix shell only opens a new shell process within the new environment, i.e., it does not change the environment of other shells, many such Nix shells can be opened in parallel. This means that Nix shells can be used as development environments.

Section 2.4 already mentions some problems with conventional development environments:

1. They are susceptible to problems faced by conventional package managers.
2. They can take up a lot of space, since each shell downloads its own, to other shells redundant components.

Problem 1. is trivially solved by Nix using the measures explained in Section 3.2. Problem 2. is also solved by Nix, since shared components necessarily have the same hash and thus occupy the same space in the Nix store.

# 4. Nix Is Broken

Although Nix solves most of the issues described in Section 2, and even yields some corollary benefits on the side, it is far from being perfect. Nix, ultimately, is a PhD project which was never intended to grow as large as it is today [39]. There are notable forks and rewrites of the Nix project which intend to fix many issues faced by Nix proper. The most notable ones are GNU Guix [40] and Snix [41]. Section 4 will explore the most prominent issues in Nix and discuss how Guix and/or Snix go about solving them, or in the case they have not been solved, what possible solutions have been put forward by the community.

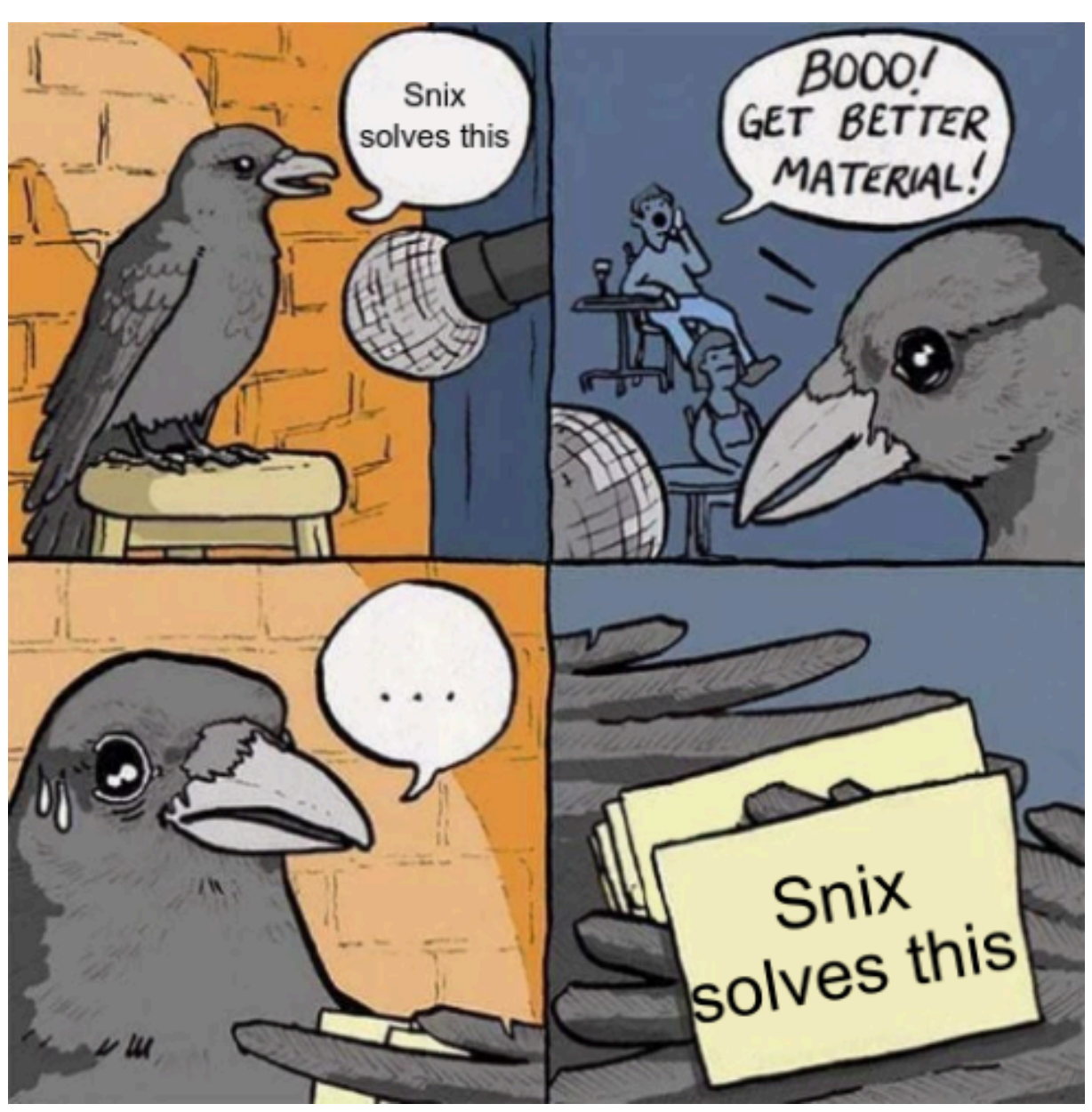

Figure 3: Snix is a rewrite of NixCPP, often cited as solving many issues faced by Nix. Image from [42].

## 4.1. Assumptions About Reference Tracking

Any automated build and deployment management system necessarily makes assumptions about build graphs and about the software being built. For instance, [3] demonstrates limitations associated with static and dynamic build graphs. In the coming sections we explore the tradeoffs associated with automated reference tracking. The concrete implementations enabled by these conditions are listed in Section A10. This is done in order to provide a later reference.

Here "reference" means a reference to a dependency in a build artifact, for example, the path to a `.so` file. In Nix references take the form of store paths.

Notation used:

- $\mathsf{dependencies}(a)$ is the set of dependencies of package $a$.

- $\mathsf{identifier}(a)$ is the string representation of reference to a package $a$.
- $\mathsf{output}(a)$ is the set of output artifacts of package $a$.
- $\mathsf{find_all}(\mathsf{identifier}(b), \mathsf{output}(a))$ returns a set of locations of reference strings found in the artifacts of $a$.

### $P_{\text{suff}}$: Reference implies dependency

If $a$ references $b$ ($\mathsf{output}(a)$ contains $\mathsf{identifier}(b)$), we may indeed conclude that a complete deployment of $a$ must include $b$. In other words, it is safe to `grep` for references. We understand this as a statement about the properties of the package naming scheme, which can be further decomposed into the following two assumptions.

#### $P_{\text{ent}}$: References can be distinguished from random data.

$$|\mathsf{find_all}(\mathsf{identifier}(b), \mathsf{output}(a))| \geq 1 \Leftrightarrow b \in \mathsf{dependencies}(a)$$

References do not occur by accident, i.e., references have a high enough entropy to distinguish them from other contents of the artifact. Nix uses a 160-bit hash as part of store paths [1], which provides enough entropy to guarantee this with a very high probability.

#### $P_{\text{inj}}$: All references are unique.

$$\forall b, c \in \mathsf{dependencies}(a)(\mathsf{identifier}(b) \neq \mathsf{identifier}(c))$$

References to two different dependencies should be distinguishable from each other. In Nix this is guaranteed by an unique hash of the store path of each package.

### $P_{\text{necc}}$: References appear verbatim at least once.

$$\forall b \in \mathsf{dependencies}(a)(|\mathsf{find_all}(\mathsf{identifier}(b), \mathsf{output}(a))| \geq 1)$$

All dependencies should have their reference appear in a findable form in the artifacts at least once. In Nix this means that each dependency has their store path appear in the artifacts in a findable format (UTF-8/ASCII) at least once [1].

### $P_{\text{rewrit}}$: References can be rewritten

The operation of substituting every occurrence of $\mathsf{identifier}(b)$ with $\mathsf{identifier}(c)$ in $\mathsf{output}(a)$ is "safe" in that it does not produce dangling references or corrupt the underlying file formats. In practice this relies on the following two criteria.

**$P_{\text{enum}}$: All references appear verbatim.**

The stronger form of $P_{\text{necc}}$, which allows to safely enumerate every occurrence of each dependency $b$ in $\mathsf{output}(a)$. In practice, this means none of the references are being obscured through, e.g., encoding or compression.

**$P_{\text{len}}$: All references have the same length.**

All references which exist in artifacts on the system have the same length. This means that references in artifacts on different computers have the same length, not just artifacts on a single system, i.e. there is a global standard reference length. This allows to exchange one reference for another without altering the structure of the artifact.

## 4.2. Trust in Local Sharing

As explained in Section 3.2.3, Nix calculates store hashes using the inputs of the derivation. This current model of the Nix store is called the *input-addressed* model, also referred to as the *extensional model* in [1]. Although Nix has worked well with IA, it causes some problems; The Nix build sandbox tries to enforce determinism, but it too can fail. For example, language non-determinism mentioned in Section 2.1.2 can cause builds to differ even if the inputs are identical. And because the store hash is calculated using the inputs (meaning source code and build instructions), this can cause two different outputs to contend over the same store path [43].

This breaks *multi-tenancy*, i.e., makes it impossible to use a shared Nix store on a machine where users distrust each other: Alice downloads a trojaned version of package $P$ from a substitutor $S_1$, which causes Alice to be compromised. Later, Bob needs the same package $P$, but does not trust $S_1$, and only trusts $S_2$. However, the Nix daemon registers $P$ as already existing in the store even though the versions served by $S_1$ and $S_2$ differ, so it is not substituted from $S_2$, and Bob has to use the trojaned version from $S_1$ [43].

### 4.2.1. Content-Addressing

The solution to the trustless shared store problem proposed by [1] is CA, also called the *intensional model.* In CA, the store hash is determined by the contents of the outputs, i.e., the build results, instead of the inputs, i.e., the build instructions [1], [43]. The problem described above ceases to exist in a CA-store, because the legitimate version and trojaned version differ in output, and thus get assigned different store hashes. Bob is shielded from the bad trust Alice has placed on $S_1$ [43].

Programs sometimes need to refer to themselves, i.e., their hosting path.[26] In the IA model this presents no issues, as the final output path of the derivation is known before building. In the CA model the final output path is not known until after the build, so if the build process needs this information, it is fed a temporary, non-existent store path $p$ with a randomly generated hash part [1].[27] After the build is finished, Nix performs *hash rewriting* on the build products. Note that hash rewriting does not guarantee a correct output binary, as the rewritten binaries may still contain references with the nonexistent path, as demonstrated in Section 4.2.3. Nix calculates the true CA path $p'$, then searches all build artifacts for references to the randomly generated path $p$ and replaces them with $p'$, then copies the files to $p'$. However, $p'$ is not calculated by content hashing over the unaltered build outputs. Instead, the hash is computed modulo self-references, i.e., the hash parts of all appearances of $p$ are replaced with an equally long string of zeroes. If this is not done, $p'$ would depend on the randomly generated $p$, which would make $p'$ non-deterministic [1].

#### 4.2.2. A Tale of Two **glibcs**

The intensional model, while solving many trust issues, presents a new, vile problem, usually called the "two glibc problem". The technical term used by Dolstra is "equivalence class collision" [1]. The user wants to compile program baz, which depends on libfoo and libbar, which both in turn depend on glibc. The local store ($L$) contains libfoo and glibc, and the substitutor ($S$) contains libbar and glibc. But because of non-determinism in the build process of glibc, the glibc versions in $L$ and $S$ differ. When the local Nix daemon substitutes libbar from $S$, it also fetches $\text{glibc}_S$, as it is a dependency. baz is now linked against two different version is of glibc, which usually makes the binary crash on execution due to duplicate symbols [1], [43], [44].

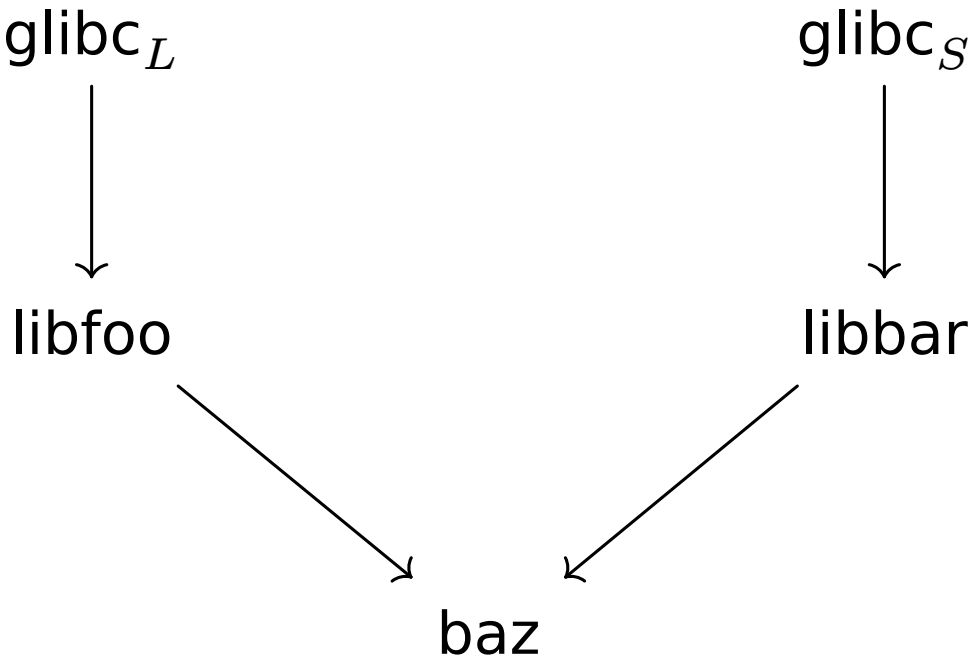

Figure 4: Equivalence class collision

[26]The final output path of the derivation can be accessed with the environment variable `$out` during the build process.

[27]While Dolstra's thesis talks about a randomly generated hash, the NixCPP implementation actually uses a hash calculated over essentially the same string as IA paths, just with a different prefix: `"nix-upstream-output:" + drvHash + ":" + drvName + optionalOutputName`, where `optionalOutputName` is empty if the output name is `out`, and `"-" + outputName` otherwise [26].

The solution to this issue is equivalence classes. An equivalence class is a hash calculated from the derivation (i.e., inputs), so it is the same hash which was used as the store hash in the extensional model. This would put $\mathsf{glibc}_L$ and $\mathsf{glibc}_S$ in the same equivalence class. To resolve equivalence class collisions Nix just makes sure that when building a derivation only one output path is selected from each equivalence class [1], [43].

So in the example above Nix would select either $\mathsf{glibc}_L$ or $\mathsf{glibc}_S$ as the sole glibc for the entire closure of baz. However, this can create potentially dangling references by violating the closure invariant, which is explained in Section A7. For example, if $\mathsf{glibc}_L$ is selected, then the dependency $\mathsf{glibc}_S$ would not be in the closure of baz despite libbar containing references to $\mathsf{glibc}_S$. To solve this issue, Nix uses hash rewriting again. A new version of libbar denoted by libbar' is created, wherein every mention of $\mathsf{glibc}_S$ is replaced by $\mathsf{glibc}_L$ [1], [43].

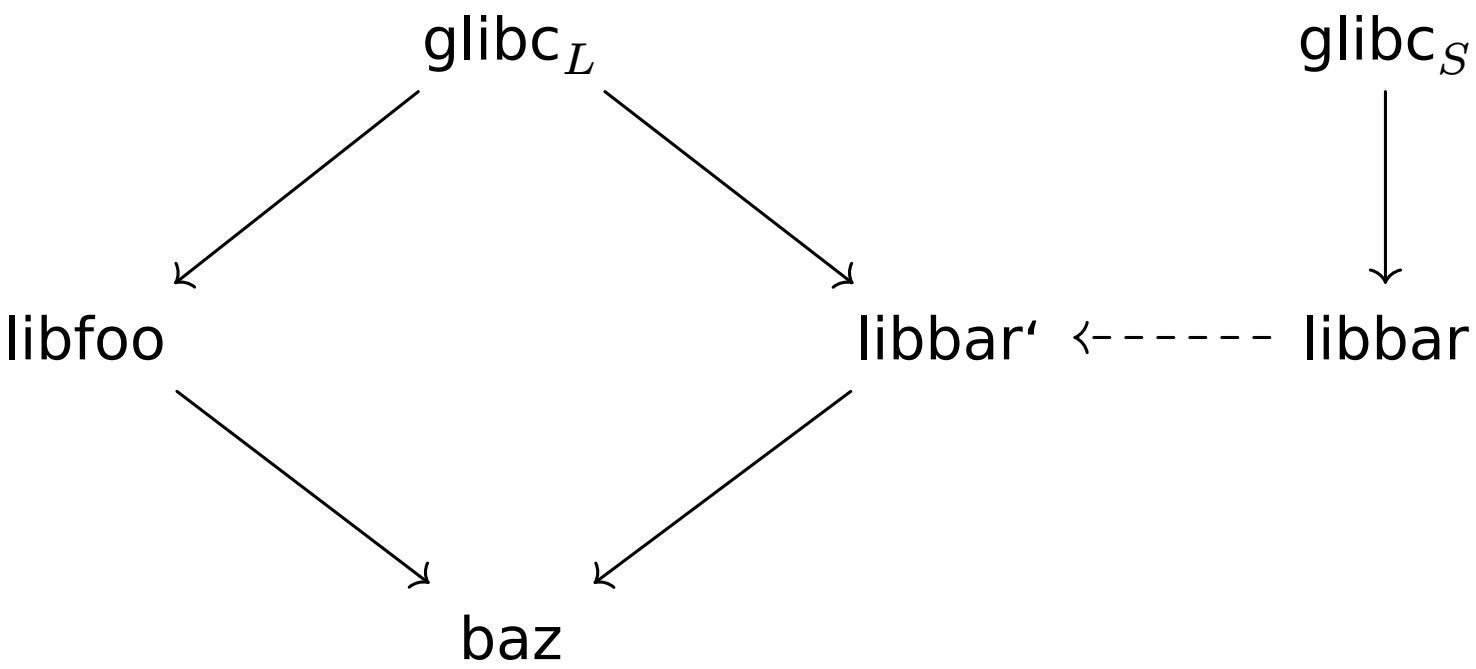

Figure 5: Equivalence class collision resolution

#### 4.2.3. Hash Rewriting

Nix guarantees $P_{\text{suff}}$, and $P_{\text{necc}}$ has been observed to not be violated in practice [1], so it can be assumed to apply. IA relies on $P_{\text{suff}} \land P_{\text{necc}}$, as an exhaustive list of runtime dependencies is needed to create the full closure of a derivation. If a runtime dependency is not found in the build artifacts, it will not be part of the resulting closure and the dependency will be removed next time the store is garbage collected.

CA requires hash rewriting. It relies on $P_{\text{suff}} \land P_{\text{enum}}$ because every reference to a dependency needs to be rewritten. It does not rely on $P_{\text{len}}$, since the entire store path is not rewritten, only the hash, which is guaranteed to be 32 characters long in Nix [1]. Since $P_{\text{necc}}$ is not guaranteed, but only assumed to be true by Nix, and $P_{\text{enum}}$ is by definition less likely to be true than $P_{\text{necc}}$, hash rewriting is harder to achieve than runtime dependency resolution. This means hash writing does not guarantee a correct output, i.e., a working binary.

#### 4.2.4. Alternatives to CA-derivations

As mentioned above, the idea of the CA-store has been around since the birth of Nix, yet the Nix infrastructure still uses IA. Mainline Nix has CA path support built in as an experimental feature [27], but Lix, a prominent Nix fork, is planning to abandon this feature altogether [45].

There is an alternative to CA which solves the trust issue of local sharing without depending on hash rewriting. This approach has been implemented in the Snix [41] project. The most important components of the Snix project are its store implementation, the `snix-store` and its storage backend, `snix-castore`. The `snix-castore` is a content-addressed data storage and syncing engine which uses a similar data storage method as Git [41].

Note that "content-addressed" here does not mean that the `snix-castore` is a CA-store, it is simply a storage backend with no knowledge of what it is storing. The store component of Snix, snix-store, uses IA, just like the NixCPP store. The "CA" in `castore` only refers to the storage method: Blobs are cut into chunks using FastCDC [46], each of which is addressable by its content hash. Deduplication is done on this chunk level. This deduplicating storage layer allows for multiple mutually untrusting users to run separate stores with separate trusted substituters, while still using storage optimally.

There is also a simpler alternative to Snix: Per-user overlay stores mounted using FUSE or virtio. Each user store would have its own Nix daemon and trusted substituters. The only shared components between users would be the base store, which contains the necessary software for booting the system. This approach does not, however, allow for deduplication like Snix does.

## 4.3. Trust in Substitution

Another form of trust in package management is the trust in the software repository, or in the case of Nix, the trust in the substitutor. The abovementioned intensional model only protects Bob against the gullibility of Alice, but it does not protect him against his own gullibility, because $S_2$ might just as well be compromised.

A solution for this provided by the Nix community is Trustix [47], a tool which compares build outputs from a set of independent substitutors to establish a sufficient degree of trust in the binaries. This solves the trust issue described in Section 2.1.2. Trustix has been implemented in the extensional model by just computing content hashes for each store path after building, but becomes easier in the intensional model, since the store hash is the content hash [47].

## 4.4. Mass Rebuilds

When a package receives a security update, applying it to the system requires a mass rebuild of all packages which depend on it, even though the only changes in their outputs are the references to the updated dependency. A more efficient method would be to reuse the already existing artefacts and only replace the references in them. There exist some solutions to this problem, most notably implemented by Guix, which allow for patching without mass rebuilds.

### 4.4.1. Grafting

If all references are unique, distinguishable, unobscured, and of same length, i.e., $P_{\text{suff}} \wedge P_{\text{enum}} \wedge P_{\text{len}}$, then true reference rewriting is possible. This allows for grafting, i.e., rewriting runtime dependencies of packages without rebuilding the package or triggering rebuilds of packages which depend on it. Grafting was added to Guix for fast deployment of critical updates [48]. With grafts, only the updated package needs to be fetched from cache or built. No other packages are rebuilt, instead their existing artifacts are searched for references to the old dependency version and have them replaced with the reference to new version. This is a much cheaper operation than a full system rebuild.

A limited form of grafting which does not rely on $P_{\text{len}}$ is currently implemented in GNU Guix; Only grafts with a store path of the same length as the replaced package are allowed [48]. Although limited in this way, it is often enough to deploy security updates. Nix has no proper grafting, but there are two resembling features. The first one is `pkgs.replaceDependencies`, which allows changing runtime dependencies of a single package without rebuilding [49]. The other one is `system.replaceDependencies`, which allows changing runtime dependencies for all packages system-wide on NixOS systems [49]. Unlike grafts in Guix, neither of these are features of the Nix package manager itself, but rather functions in nixpkgs and NixOS respectively. This prohibits system-wide grafting on non-NixOS systems.

### 4.4.2. Early and Late Binding

The "binding time" in Nix informally refers to the moment the presence of a dependency in the runtime closure is determined, e.g., a reference to a concrete path to a dependency is embedded in any output. In nixpkgs, it is conventional to link dependencies early: An executable is usually "bound" to its shared libraries in the same derivation it is built in during the `fixupPhase`.

Late binding allows packages to be built against the oldest library possible and the appropriate version is loaded at runtime [50]. This feature would enable deploying security updates at runtime, i.e., without rebuilding, and in addition it would also make some development tasks easier; For example, when developing software in a dev env with nixpkgs pinned to a newer version than the system nixpkgs, the libc versions might mismatch. Instead of crashing due to not finding the appropriate libc, late binding would allow loading the system libc at runtime.

Enforcing early binding can sometimes result in problems with the software itself; Many drivers rely on late binding [50], as does the Python ecosystem. On conventional Linux systems Python packages use late binding, whereas in nixpkgs they are forced to use early binding: References to their dependencies are embedded in `${!outputDev}/nix/propagated-build-inputs`, which are then pulled into the final closure of the `python3.withPackages (...)` wrapper [51]. However, there is an out-of-tree (i.e., not in nixpkgs) solution for this called Pyproject.nix [52].

## 4.5. `stat` Storm

When an ELF built with Nix or Guix is ran, the dynamic loader searches all directories listed in the `RUNPATH` entries of the ELF for the required libraries [53]. This creates what is known as a `stat`-storm.[28]

A solution to the `stat` storm is `ld.so.cache`, i.e., the dynamic linker cache file. In conventional Linux systems this file, residing at `/etc/ld.so.cache`, serves as a system-wide cache, mapping shared library names to their filesystem locations [53]. It allows the dynamic linker to find libraries directly by their full path, without having to search each directory. Guix has successfully used a local `ld.so.cache` to act as a per-application loader cache [55]. The local `ld.so.cache` is populated with all the filesystem locations of the dependencies, and the ELF is configured to prefer the cache over `RUNPATH` entries.

## 4.6. Relocatability

A feature that would be made possible by reference rewriting is relocatability, i.e., the ability to modify binaries to use dependencies from a store in a different location. When moving binaries to a store in another location in the filesystem, either on the same host or on another host, all references have to be rewritten to use the new store location without further changing the structure of the artifacts. This operation won't change hashes, since every time file contents with a store path are used to calculate a hash, the store location can be replaced by a placeholder in a similar manner such as in the intensional model the temporary output path hashes are replaced by zeroes [1].

This kind of relocatability is not yet implemented in Nix, although there have been efforts [56], [57]. However, Guix has a substitute in the form of relocatable software bundles which can be run even on systems without Guix installed. It works by creating a wrapper[29] around binaries which populates `/gnu/store` in the mount namespace of the process. In addition, an environment has to be entered before running the binary

[28] `stat(2)` is the Linux system call for getting the status of a file [54].

[29] This wrapper is a statically linked C program, so it does not have dependencies.

to set variables such as `GUILE_LOAD_PATH` [58]. A similar solution has been proposed for Nix [59].

## 4.7. Impurity in the Builder

The Nix build sandbox performs very well; Malka, Zacchiroli, and Zimmermann [60] report 99.99% reproducibility of build environments using a set of over seven million packages coming from 200 different revisions of nixpkgs. Note that a build environment contains all the components and variables necessary to build the software, but it does not include the software itself. They conclude that Nix is able to achieve perfect build environment reproducibility, since the 0.01% of irreproducible builds stem from using Nix features which cause impure evaluation by design.

Another paper by the same authors [61] investigates the bitwise reproducibility rates of the build outputs of over 700 000 nixpkgs packages, taken from revisions between 2017 and 2023. The result is a reproducibility rate between 69 and 91% with an upward trend. Note that this study used shallow builds, i.e., build environments were downloaded from the NixOS cache instead of being built from scratch. However, considering the 99.99% reproducibility rate of build environments established in the other paper, full builds would not change the results significantly. Reasons for irreproducibility were build timestamps (14.8%), embedded environment variables (2.2%), embedded build IDs[30] (2.2%), `uname` outputs (1.3%). A total of 19.7% of irreproducible packages were identified to be caused by one of these four categories. These leaks of random data into the build environment happen, because, although it is well isolated for the most part, there are still some sources of randomness, e.g., the system clock and some system info such as the kernel version, that are still accessible inside the sandbox.

Malka, Zacchiroli, and Zimmermann [61] recommend the following actions to be taken for mitigating irreproducibility: Builders should publish signed attestations and artifact checksums, and Nix should have an inbuilt mechanism which verifies that build artifacts match the consensus checksum. These precautions do not solve irreproducible derivations per se, but help in detecting them. When such a derivation is found, the reproducibility problem will have to be fixed manually on a case-by-case basis.

## 4.8. Incremental Builds

Reusing artifacts from earlier builds is called incremental building [62]. Only the parts of the program which are affected by the change are rebuilt, thus decreasing the time spent building, and thereby speeding up development. According to Maudoux and Mens [63], incremental builds can improve build performance by up to two orders of magnitude. Unfortunately, Nix only supports incremental builds on the granularity of packages [64]. On the level of modules[31] Nix does not perform any incremental builds. If any part of the source code is changed, the entire package is rebuilt from scratch.

[30] A form of language non-determinism; The compiler embeds a unique and nondeterministic ID into the artifact.

The Garnix project [65] has developed a rudimentary substitute for incremental builds [64]. Adding an output with the name `intermediates` to the derivation allows future builds to access the contents of this output by referring to `cache`. For each build, the Garnix CI checks if the past five builds have succeeded, and exposes the `intermediate` output of the most recent one as the `cache` of the current build. If none of the five most recent builds have succeeded, a clean, non-incremental build is ran. Garnix itself does not check whether artifacts in `cache` are stale, the burden is instead placed on the developer.

There are two major issues with this approach [64]:

1. It is not easy to find out which derivations need to be built on a local system to match the state of the cache.
2. There might be unnecessary rebuilding on the Nix package level due to the cache changing.

The Garnix project maintains that both will be solved in the future [64].

Complementary approaches to solving this issue include subdividing Nix packages into multiple smaller packages on the required level of granularity. However, doing this requires some advanced, experimental, and highly controversial Nix features, such as dynamic derivations and import from derivation (IFD), which are discussed in Section 4.9. In addition, building multiple small Nix packages adds a substantial amount of overhead, which defeats the point of incremental builds [64].

There also exist alternative build systems with more mature support for incremental builds. An example is Bazel, developed by Google, based on their internal Blaze build tool [66]. Bazel supports granularity on the level of modules and reusing outputs from cache based on their content hash [66].

## 4.9. Dynamism

Nix encourages a separation of build planning, i.e., constructing the build graph, and build execution [67]. Build planning is handled by the Nix language, which produces derivations, which in turn are used to execute the building. Dolstra [1] describes this setup as desirable feature due to its conceptual simplicity. This type of build system where the build plan is statically known is called *applicative*, because it behaves like

[31] An example of a module is a single object file.

the `apply` or `<*>` function of the applicative functor [3], [34].[32]

```haskell
apply :: Applicative f => f (a -> b) -> f a -> f b
```

Listing 5: The type signature of the `apply` or `<*>` function written in Haskell [34].

The applicative build system is isomorphic to the `apply` function depicted in Listing 5. The first input of the function is `f (a -> b)`, an action which defines how to build `b` given `a`. The second input `f a` is an action which builds `a`. The output is `f b` an action which builds `b`. This means that the output of action `f a`, namely `a`, cannot influence `f b`, i.e., the build plan for `b`. All parts of the build plan need to be known before any builds can be ran [34].

```haskell
bind :: Monad m => m a -> (a -> m b) -> m b
```

Listing 6: The type signature of the `bind` or `>>=` function written in Haskell [34].

In contrast, a *monadic* build system needs to run builds to construct the build graph [34]. The monadic build system is isomorphic to the `bind` function as depicted in Listing 6. The first input of the function is `m a`, i.e., an action which returns `a`. The second input is `a -> m b`, a function which generates `m b`, i.e., an action which generates `b`. The output is `m b`. This means that the function `a -> m b` uses the result of `m a` to generate `m b`, i.e., `m a` is executed to get its build result, `a`, which then determines a new build recipe for `b`, i.e., `m b`. More generally, to know the next step in the build plan, i.e., `m b`, the build system needs to execute an earlier build, i.e., `m a`. This execution of builds to determine other parts of the build graph is not possible in the applicative model [34].

Dynamism in Nix refers to different methods of enabling monadic building. Nix allows dynamism in the form of IFD [27], but it is restricted to only applicative actions in nixpkgs [34].[33] IFD is when the value of a Nix expression can depend on the contents of a store object [27].

```nix
let
  pkgs = import <nixpkgs> {};
  inner = pkgs.runCommand "inner" {} ''
    sleep 10;
    echo "Hello from inner." > $out
  '';
  ifd_inner = builtins.readFile inner;
in
  pkgs.runCommand "outer" {} ''
    echo ${ifd_inner} > $out
    echo "Hello from outer" >> $out
  ''
```

Listing 7: A Nix expression using IFD, adapted from [68].

[32] Although notation and jargon from category theory is used in this section, it is not required to understand them in order to understand the difference between applicative and monadic build systems.

[33] This is because Hydra uses the `restrict-eval`-setting, which prohibits use of IFD [34].

Listing 7 shows a practical example of IFD. Building the package for the first time takes 10 seconds, as expected [68]. If the line `sleep 10` is changed to `sleep 20`, the content of the outer Nix expression does not change, and the build output does not change. Despite this, building again takes 20 seconds, since the IA hash of inner Nix expression has changed, and Nix has no way of knowing whether that change will affect the ingested value of the outer Nix expression before realizing the inner derivation. Building again with no change is fast because there has been no change to the inner expression, and thus Nix knows that there cannot possibly be a change to the outer expression either [69]. The Nix evaluator implemented in NixCPP[34] cannot resume evaluating while waiting for a build [34]. This pausing of evaluation is the reason IFD has been banned from nixpkgs [68]. However, the community has built various projects to leverage IFD in parsing lock files and to bring incrementality to Nix. These projects are collectively called `lang2nix` [68].

```
let
  pkgs = import <nixpkgs> {};
  producing =
    pkgs.runCommand "inner.drv" {
      outputHashMode = "text";
      requiredSystemFeatures = ["recursive-nix"];
    } ''
      echo "let pkgs = import \"${pkgs.path}\" {};
            in
            pkgs.runCommand \"inner\" {} '''
              sleep 10;
              echo \"Hello from inner!\" > \$out
            '''
            " > inner.nix
      cp $(${pkgs.nix}/bin/nix-instantiate inner.nix) $out
    '';
in (builtins.outputOf producing.outPath "out")
```

Listing 8: Listing 7 rewritten using dynamic derivations, adapted from [68].

An alternative to IFD are dynamic derivations. This feature allows Nix to create additional derivations at build time to expand the build graph [68]. This removes the blocking issue present in IFD [34]. Practical examples of using dynamic derivations in practice are the MakeNix program by Farid Zakaria [70], [71] and nix-ninja [72]. It adds incrementality to C/C++ builds by using the `-MM` option of GCC and Clang, which emits depfiles that contain Makefile targets with dependency information. These depfiles are parsed by a Go program which transforms them into a Nix expression. Each object file is generated by its own derivation, and when the source code is changed, only the affected object files get rebuilt.

[34] Snix is not affected by the non-parallel evaluation present in IFD [34].

Dynamic derivations can be enabled in Nix as an experimental feature [68]. Due to its experimental nature and recent addition to Nix it has not been allowed in nixpkgs and has also been used less in out-of-tree projects than IFD. However, as dynamic derivations do not suffer from the blocking issue as IFD does, it may have better chances of being adopted into nixpkgs in the future.

# 5. Findings

Nix has been able to solve many software deployment problems with its functional approach. Requiring explicit inputs has greatly improved repeatability and reproducibility of software builds. The most common dependency issues have been solved by the introduction of hashes. The problems of CCMS have been solved by NixOS with the congruent model, and Nix shells have greatly improved development environments.

Despite the success of Nix, some of the problems remain unsolved. In addition, the functional model has generated some new problems, too. These problems, their proposed solutions, and workarounds are listed in Table 1 below.

| **Issue** | **Solution** | **Alternative** |
|---|---|---|
| Trust in local sharing (multi-user stores) | CA-derivations | Per-user stores (Snix) |
| Trust in substitution | Trustix | |
| Mass rebuilds | Grafting, late binding | |
| `stat` storm | `ld.so.cache` | |
| Relocatability | Reference rewriting | Portable software bundles, relative references |
| Impurity, irreproducibility | Sandbox improvements, consensus checksums for detection of irreproducibility | Repairing packages manually to be more reproducible. |
| Incremental builds | Finer granularity, i.e., smaller derivations. | Dynamic build graphs |

Table 1: Remaining problems in Nix, their proposed solutions, and workarounds.

Trust is mainly unsolved in current Nix but the technical solutions exist. For local sharing this solution is the CA-store. However, as was discussed earlier, CA derivations have strict requirements which are hard to guarantee. An alternative to CA-stores are per-user traditional IA-stores. This approach is implemented successfully and with minimal drawbacks by Snix. Trust in substituters can be solved by Trustix, however, it has not been implemented widely.

Mass rebuilds as a consequence of security updates or other patching has been solved by Guix with grafting. Furthermore, slow startup speeds have been mitigated by using a local `ld.so.cache`. Similarly, relocatable binaries have been partly solved by Guix in the form of portable software bundles.

The state of the Nix sandbox has been improving significantly over the past years, leading to very good reproducibility figures. There have not been any major innovations in adjacent projects regarding the sandbox.

Incremental builds are something Nix, Guix, Snix, Lix, etc. are struggling with. While workarounds using dynamic build graphs (i.e., what [3] describes as monadic build systems) exist, they do not grant proper module level granularity and are not allowed in nixpkgs.

# 6. Discussion

The findings demonstrate that the functional model of Nix has been successful in addressing the main issues of software deployment. In addition, the remaining problems have received significant attention from the community, especially during the recent years. Trust in local sharing, security updates, relocatability, and `stat` storms have all been successfully solved, albeit partly with workarounds. The issue of trust in substitution has been solved in theory by Trustix. Furthermore, the constantly improving reproducibility figures of nixpkgs indicate that the problem of builder impurity is handled well by the community. To summarize, solutions and workarounds exist, but their adoption into mainline Nix and nixpkgs is lagging behind.

One of the truly unsolved issues is incrementality, one of the biggest weak points of Nix. Dolstra chose separation of planning and building as a design feature of Nix in its early stages, so forcing dynamic build graphs into Nix requires working against its own design in some sense. Despite this, IFD and dynamic derivations have been implemented in Nix, and are used by numerous out-of-tree projects. In [34] Lovelace introduces the notion of the postmodern build system, a hypothetical future build system with many desirable features. A postmodern build system has trustworthy incremental builds, reuse of compilation, and distributed building. They conclude that Nix is not a postmodern build system, but the idea of making `execve` pure, i.e., derivations, is a necessary feature for a hypothetical future postmodern build system. The inability to perform true module-level incremental builds is not an inherent flaw of the functional approach, but rather a design issue of Nix, and consequently Guix, Snix, and Lix too.

It should be noted that the solutions to both trust in local sharing, i.e., multi-user stores, and irreproducibility involve content hashing. In the first one in the form of CA-derivations and in the latter one in the form of consensus checksums. This goes to show that there is overlap between solutions, and a solution for one problem may yield improvements in other areas too.

Despite the broad scope of this thesis, there are still many topics which could be explored. Left unaddressed in this overview of Nix were, e.g., scheduling and resource allocation aspects [73]–[75], minimal bootstrap seed [76] and supply chain security [77].

# 7. Conclusion

It is clear that the benefits of the purely functional approach outweigh its costs. The momentum of the Nix, Snix, and Guix communities has proven that they are viable solutions, not just experimental toys. In the future, the Nix community should work on porting working solutions, such as grafting, portable bundles, and local `ld.so.cache` from Guix to mainline Nix. Additionally, Trustix-style verification should be incorporated into nixpkgs infrastructure. As Snix matures, its modular approach should take NixCPP's place as the default implementation, as the content-addressed storage approach of `snix-castore` yields significant benefits. However, there is still the issue of granular incremental builds, which both Nix and Nix-adjacent projects such as Guix have been unable to solve. This may warrant the creation of an entirely new tool in the future, incorporating features from both Nix and Bazel. Despite this, the purely functional approach is the way forward for software deployment.

# A Appendix

## A1 Nix Package Manager Concept Map

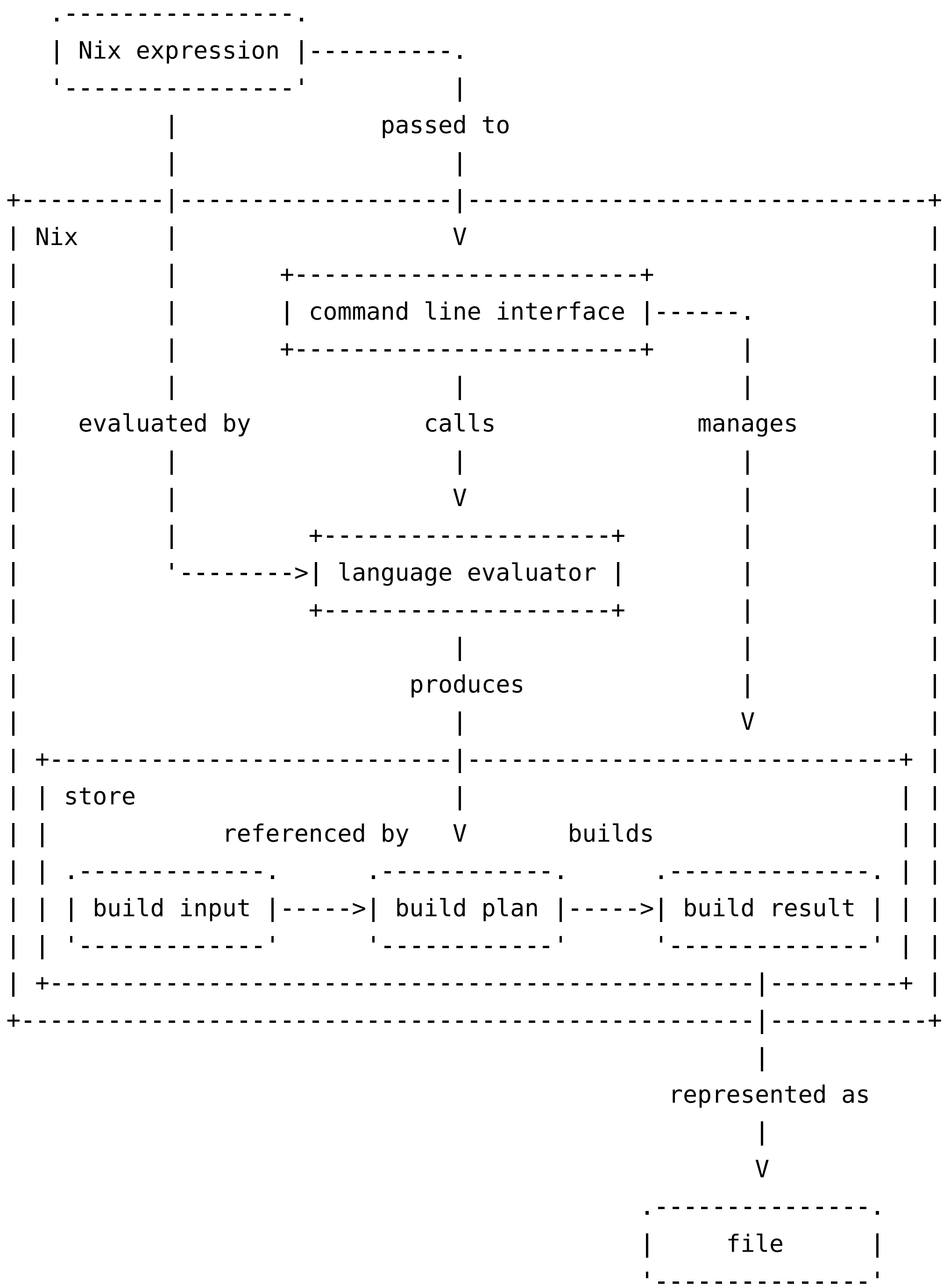

Figure 6: Nix package manager concept map from the Nix reference manual [27].

## A2 The Definition of "Functional"

Not only is the Nix DSL functional, but the model of software deployment that Nix follows is called "functional". This is because in Nix, the artifacts which end up in the store depend entirely on the inputs.[35] This is similar to how in functional programming the result of a function depends purely on its inputs. In both Nix and functional programming, there are no side effects [1].

## A3 Store Path Construction

E. Dolstra [1] and the Nix manual [27] specify how an IA store path is to be constructed:

$$\text{storePath} = \text{nixStore} + \texttt{"/"} + \textsf{truncatedHash}(\text{fingerprint}) + \texttt{"-"} + \text{name}$$

where

$$\begin{aligned}&\text{fingerprint} = \\ &\quad \text{type} + \texttt{":sha256:"} + \textsf{truncatedHash}(\text{innerFingerprint}) + \texttt{":"} + \text{nixStore} + \texttt{":"} + \text{name}\end{aligned}$$

- nixStore is the path of the Nix store on the filesystem (usually `/nix/store`).
- truncatedHash is a function for printing the truncated hash of a string. See Section A4.
- name is the name of the output.
  Because a single derivation can have multiple outputs,[36] and each one must have an unique name, the name includes a postfix for each output, such as `lib` or `doc`. For the main output (`out`), the postfix is omitted. In praxis, the version of the packaged software is also included in the name [27].
- type is the type of store object.
  Either text, source, or output. Text is used for derivation files, source for the source FSO (i.e. source tree), and output for derivation outputs.

innerFingerprint is one of the following, based on type [27]:

- Text: The contents of the resulting store path, e.g., the ATerm string inside a `.drv` file.
- Source: The Nix archive (NAR) serialization of the source FSO. See Section A8 for a more technical explanation.
- Output: The ATerm string of the derivation, with the output path(s)[37] set to an empty string.[38]

[35] Or rather, this is the goal. As Section 4.7 demonstrates, Nix still has some minor issues regarding build process purity.

[36] For example: A single program provides both a runnable binary and libraries associated with it. Most users only want the runnable binary, so it is set as the main output of the Nix package, and the libraries are set as a secondary output.

[37] Recall that a derivation can have multiple outputs.

## A4 Hash Truncation

The `truncatedHash` function mentioned in Section A3 calculates the SHA-256 hash of the input, which is then truncated to 160 bits [1], [27].[39] The reason for the truncation is in an early design decision and undesirable coding practices.

In the past Nix used 128-bit MD5 hashes, which consist of 32 characters. However, MD5 became unsafe shortly after the Nix project began, and thus a move to SHA-1 was required [78]. SHA-1 produces a 160-bit hash, but switching to a base32 representation kept the hash at 32 characters[40] and ensured backwards compatibility. This compatibility was required because some builders have the 32-character hash length hard-coded [1].

In 2017 SHA-1 became broken, and thus a move to SHA-256 was warranted [79]. However, SHA-256 hashes are (as the name suggests) 256 bits long, and as such would break the backwards compatibility of 32-character long hashes. Some filesystems, e.g., ISO-9660 with the Joliet extension[41] allow only 64 characters long filenames, and thus the decision was made to truncate the SHA-256 hash to 160 bits and retain the 32-character long hash part [1].

## A5 Garbage Collector Roots

Garbage collector roots, or gcroots for short, are store paths (and by extension their dependencies) which will not be removed from the store by the garbage collector. An example of a gcroot is the current generation in a NixOS system. Any store object which has a symlink in `/nix/var/nix/gcroots` or a subdirectory contained within it is considered to be a gcroot [27].

## A6 Closures

The Nix package manager has full knowledge of all dependencies of a derivation. The full tree of dependencies of a single object is called its closure. The formal definition by Dolstra: The closure of a set of files $C : \{(\mathsf{Path}, \mathsf{Content})\}$ is the smallest set $C' \supseteq C$ satisfying

$$\forall (p, c) \in C' : \forall p_{\text{ref}} \in \mathsf{references}(c) : \exists (p', c') \in C' : p_{\text{ref}} = p'$$

[38]Restricting the hash to be calculated only over the inputs prevents a causality dilemma, since the output path, which contains said hash, is also part of the derivation.

[39]By default, Nix uses SHA-256 hashes, however, other hash types are also supported [27].

[40] $\frac{160}{\lg(32)} = 32$

[41]Nix needed to support the ISO-9660 filesystem since historically system installers were booted from CD-ROMs.

where the tuple $(p, c)$ denotes a path $p$ and the contents $c$ required at $p$, and the relation $\mathsf{references}(c)$ denotes the set of references contained in the content $c$. [1]

## A7 The Closure Invariant

The closure invariant is an invariant of the Nix store, which states that for a path in the Nix store to be valid, it needs to be closed under the references relation. The references relation is nothing more than a mapping from a store path to the set of other paths it references: $\mathsf{references} : \mathsf{Path} \rightarrow \{\mathsf{Path}\}$. For a path $p$ to be closed under this relation means that all references of $p$ are valid too, or more formally:

$$\forall p \in \mathsf{Path} : (\mathsf{valid}[p] \neq \varepsilon \Rightarrow \forall p' \in \mathsf{references}[p] : \mathsf{valid}[p'] \neq \varepsilon)$$

where the expression $\mathsf{valid}[p] \neq \varepsilon$ means that the store path $p$ is valid [1]. In effect this invariant guarantees that there can exist no dangling references in the Nix store.

## A8 The Nix Archive Format

The NAR format is a data serialization format designed to serialize FSOs (directories, files, and symlinks) [1]. There already exists a plethora of FSO serialization formats, such as TAR or ZIP, but they can behave non-deterministically. NAR behaves much like TAR, but guarantees deterministic output, that is, there exists a bijective function from each FSO to a NAR [27].

## A9 The NixOS Boot Process

The NixOS boot process is divided into three parts; First the bootloader loads the initial ramdisk, which contains the kernel and necessary software to mount the root filesystem later. The kernel starts the stage 1 init script, which mounts the root filesystem. The stage 2 init script, which is located in the Nix store on the root filesystem, is started by the stage 1 script. This script performs all remaining tasks of the boot process [36].

## A10 Table of propositions and implementations

| **Implementation** | $P_{\text{suff}}$ | $P_{\text{necc}}$ | $P_{\text{enum}}$ | $P_{\text{len}}$ |
|---|---|---|---|---|
| IA-derivations | ✓ | ✓ | | |
| CA-derivations | ✓ | ✓ | ✓ | ✓ |
| Relocatability | ✓ | ✓ | ✓ | ✓ |
| Grafting | ✓ | ✓ | ✓ | ✓ |

Table 2: Implementations and their required assumptions.